\documentclass[preprint,12pt]{elsarticle}

 \usepackage{graphicx}

\usepackage{amssymb}

\biboptions{comma,square,sort&compress}

\journal{Optics Communications}

\begin{document}

\begin{frontmatter}


\title{Entanglement dynamics of two-bipartite system under the influence of dissipative environments}
\author{Smail Bougouffa}
\ead{sbougouffa@hotmail.com}
\address{Department of Physics
Faculty of Science, Taibah University, P.O. Box 30002, Madinah,
Saudi Arabia\fnref{label3}}

\begin{abstract}
An experimental scheme is suggested that permits a direct measure
of entanglement of two-qubit cavity system. It is articulated on
the cavity-QED technology utilizing atoms as flying qubits. With
this scheme we generate two different measures of entanglement
namely logarithmic negativity and concurrence. The phenomenon of
sudden death entanglement (ESD) in a bipartite system subjected to
dissipative environment will be examined.
\end{abstract}


\end{frontmatter}

\section{Introduction}
Entanglement is an essential feature of quantum mechanics that
permits basic peculiarity between classical and quantum physics.
It is at the heart of many applications of quantum information
science, including quantum teleportation
\cite{BBCJPW93,BPMEWZ97,BBDHP98, FSBFKP98},quantum dense coding
\cite{BW92}, quantum cryptography \cite{E91}, and quantum
computing \cite{BR01}. Entanglement can reveal the nature of a
nonlocal correlation between quantum systems that have no
classical interpretation. Recently, the use of the cavity QED in
quantum information processing becomes more interesting
\cite{ECZ97,R99,OM00,JPJ02,ZG01}, and the entanglement generation
and nonlocality test of two cavity fields have been explored
\cite{MKB04,BP03,ILZ07,TIAZ08,LX04}. Real quantum system are
unavoidably subjected to their environments, and these reciprocal
interactions often result in the dissipative evolution of quantum
coherence and loss of useful entanglement. Decoherence may guide
to both local and global dynamics, which may incite the eventual
deterioration of entanglement \cite{ILZ07}. Particularly, Yu and
Eberly have shown that the entanglement of bipartite systems can
decay to zero abruptly, in a finite time which depends upon the
initial preparation of the atoms, a phenomenon termed entanglement
sudden death (ESD)\cite{YE07} and was recently observed in two
sophistically designed experiments with photonic qubits
\cite{A07}and atomic band \cite{L07}. Furthermore, it has also
been observed in cavity QED and trapped ion systems \cite{S06}. On
the other hand, the phenomenon ESD has provoked many theoretical
investigations in other bipartite systems involving pairs of
atomic, photonic, and spin qubits \cite{MLK08, G08, TPA05, CYH08},
multipartite systems \cite{LRLSR08} and spin chains
\cite{CP08,LHMC08,AGXWL06}. In addition, ESD has also been
explored for different environments \cite{ TIAZ08, YE07,FT06,
FT08}. However, numerous investigations on ESD in a variety of
systems have been done so far, the question of ESD in interacting
qubits
remains yet open \cite{DA09}. \\
On the other hand, several methods \cite{SMRMZ06, GHBELMS02, HE02,
H03, CZG06} have been proposed for detecting and measuring
entanglement without a full reconstruction of the state. These
methods, although much simpler than the full state reconstruction,
are not completely free of experimental difficulties, as they
require either controlled unitary operations or some prior
knowledge about the quantum state in question, or they can detect
entanglement but not measure its amount. Furthermore, for two
qubits the defining measure of entanglement is concurrence
\cite{W98}. It is a good measure of entanglement in every sense
and direct measure of the concurrence of two-photon pure entangled
state was confirmed experimentally using linear optical means
\cite{WSDMB06}. Nevertheless, it exists another computable measure
of the entanglement called negativity \cite{VW02}, and thereby
fill an important gap in the study of entanglement.  It can be
regarded as a quantitative version of Peres' criterion for
separability \cite{P96}. For a mixed quantum states,
the two measures are different. We will show some analytic relations between
the two previous measures of entanglement for the proposed system.\\
The aim of this paper is two-fold. First, we propose an efficient
scheme for quantum teleportation to generate entangled number
states of two-bipartite system under the influence of dissipative
environments. Second, we present two different computable measures
of entanglement namely, the logarithmic negativity and the
concurrence and we compare their amounts and agreements for
different reservoirs. Thus we investigate the problem of ESD for a
this proposed scheme.\\
The format of this paper is as follows. In section \ref{sec2} we
investigate the model for two-bipartite system with a simple
dissipative reservoir and formulate their dynamical evolution by
solving the master equation of motion. In section \ref{sec3} we
present the theory of two different measure of entanglement of
two-bipartite system subjected to dissipative environments (the
logarithmic negativity and the concurrence). In section \ref{sec4}
we study the entanglement dynamics of two-bipartite system in
vacuum and thermal reservoirs. Then we compare the amounts of the
two computable measures of entanglement for different initial
entangled states. We find that for thermal reservoirs the ESD
always exists but for the vacuum reservoirs the ESD can be exhibit
with some entangled states. Finally in section \ref{sec5} we
conclude with a general remarks and future outlook.

\section{Model}\label{sec2}
\begin{figure}
  \center{
  \includegraphics[height=0.4\textheight]{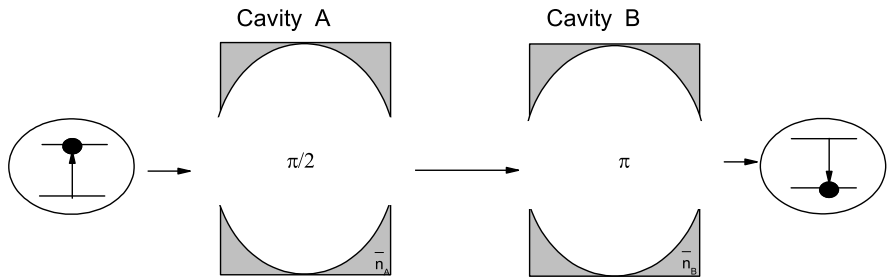}}
  \caption{The scheme for single-particle interference.
  A two-level atom prepared in its excited state passes successively through cavity A,
  a field damping region and cavity B. At the end the two-level atom will be in its ground
  state.}\label{Fig0}
\end{figure}

Recently, Zubairy et all \cite{ZAS04} have proposed a new scheme
in their investigation of the quantum disentanglement eraser. In
this simple scheme , the concurrence can be directly measured from
the visibility for an explicit class of entangle states. We
propose here the same scheme but with some modifications. A
two-level atom with the upper level $|e\rangle$ and the lower
level $|g\rangle$ passes consecutively through cavity A, a field
damping region and a cavity B as shown in figure \ref{Fig0}. The
atom is initially prepared in the excited state $|e\rangle$ and
the decay of the radiation field inside a cavity may be described
by a model in which the mode of the field of interest is coupled
to a whole set of reservoir modes. We assume that initially the
two cavities are in vacuum state $|0\rangle$ and the atom always
leaves the setup in the ground state $|g\rangle$. \\
In the interaction picture and the rotating-wave approximation,
the Hamiltonian is simply
\begin{eqnarray}\label{1}
  H &=& \hbar\sum_{j}\left[ g_{j}^{(A)} b_{j}^{(A)\dag}a_{A}
  e^{-i(\nu-\nu_{j})t}+ g_{j}^{*(A)}a_{A}^{\dag}b_{j}^{(A)}
    e^{i(\nu-\nu_{j})t}\right] {}   \nonumber\\
    & & {} + \hbar\sum_{j} \left[ g_{j}^{(B)} b_{j}^{(B)\dag}a_{B}
e^{-i(\nu-\nu_{j})t}+g_{j}^{*(B)}a_{B}^{\dag}b_{j}^{(B)}
    e^{i(\nu-\nu_{j})t}\right]
\end{eqnarray}
where $a_{A(B)}$ and $a^{\dag}_{A(B)}$ are the destruction and
creation operators of the mode of the electromagnetic field of
frequency $\nu$. $b_{j}^{A(B)}$ and $b_{j}^{\dag A(B)}$ are the
modes of cavity A(B) of frequency $\nu_{j}$ which damp the field
and $g_{j}^{(i)}$ is the coupling constant of the interaction
between the electromagnetic field and the cavity.\\
From the general analysis of system-reservoir interactions, with
the Hamiltonian (\ref{1}), we can obtain directly the master
equation for the reduced density matrix for the filed in the
cavities as \cite{SZ97}
\begin{eqnarray}\label{2}
  \dot{\rho}(t) &=& -\sum_{i=A,B}\frac{\gamma^{(i)}}{2}(\overline{n}_{i}+1)\left[a_{i}^{\dag}a_{i}\rho(t)-
  2a_{i}\rho(t)a_{i}^{\dag}+\rho(t)a_{i}^{\dag}a_{A}
  \right]\nonumber\\
   &&-\sum_{i=A,B}\frac{\gamma^{(i)}}{2}\overline{n}_{i}\left[a_{i}a_{i}^{\dag}\rho(t)-2a_{i}^{\dag}\rho(t)a_{i}+
   \rho(t)a_{i}a_{i}^{\dag}
   \right]
\end{eqnarray}
where $\gamma^{(i)}$ is the decay rate in the cavity, and
${\overline{n}}_{i}(i=A,B)$ are the average number of quanta at
frequency $\nu$ in the thermal reservoir which surrounds the
cavities A and B. If the reservoirs are at zero temperature,
$\overline{n}_{i}=0$, and the remaining terms are due to vacuum
fluctuations. \\
In the general case, we consider the field states in Fock basis in
two identical high-Q cavities A and B that represent a bipartite
system containing the entangle field as
\begin{equation}\label{19}
    |\Psi\rangle_{AB}(0)=a_{1}|0\rangle_{A}|0\rangle_{B}+a_{2}|0\rangle_{A}|1\rangle_{B}+
    a_{3}|1\rangle_{A}|0\rangle_{B}+a_{4}|1\rangle_{A}|1\rangle_{B}
\end{equation}
where $a_{i} (i=1,2,3,4)$ are the probability amplitudes with
$\sum_{i=1}^{4}|a_{i}|^{2}=1$. We use the basis (
$|1\rangle=|0\rangle_{A}|0\rangle_{B},|2\rangle=|0\rangle_{A}|1\rangle_{B},
|3\rangle=|1\rangle_{A}|0\rangle_{B},|4\rangle=|1\rangle_{A}|1\rangle_{B}$)
to define the density matrix of the two qubit system. The
equations of motion in terms density matrix elements can be
obtained using the master equation \ref{2} and with their
solutions in the general case are given in the Appendix A.

\section{Degree of entanglement}\label{sec3}
\begin{figure}
  \center{
  \includegraphics[height=0.3\textheight]{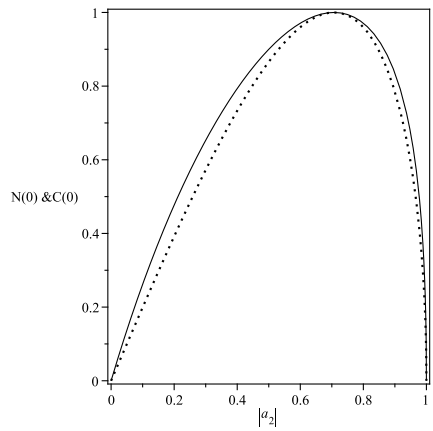}}
  \caption{Initial value of logarithmic negativity (line) and Concurrence (dot) in terms of $|a_{2}|
  $ where $|a_{2}|^{2}+|a_{3}|^{2}=1$}\label{Fig1}
\end{figure}

To study the effect of interaction among the two-bipartite on
decoherence we have to investigate the dynamics of two-bipartite
entanglement. In order to compare the degree of the entanglement
restrained to quantum state, we will use two entanglement measures
, i.e., logarithmic negativity and concurrence, to describe the
degree of entanglement for any bipartite system. Both measures
satisfy necessary conditions for being good measures of
entanglement. The logarithmic negativity \cite{VW02, LCOK08} for
two-bipartite system is defined by
\begin{equation}\label{13}
    \mathcal{N}=\log_{2}\|\rho^{T_{B}}\|_{1}
\end{equation}
where $\rho^{T_{B}}$ is the partial transpose of a state $\rho$ in
$d\otimes d'$ ($d\leq d'$) quantum system and $\|.\|_{1}$ is the
trace norm that can be read as
\begin{equation}\label{14}
    \|\rho^{T_{B}}\|_{1}=1+2|\sum_{i}\mu_{i}|
\end{equation}
where $\mu_{i}$ are the negative eigenvalues of $\rho^{T_{B}}$. For pure states,
$\mathcal{N}=0$ for unentangled states and $\mathcal{N}=0$ for the maximally entangled state. \\
We will also consider another important measure of entanglement
that is the concurrence \cite{W98, W06},
\begin{equation}\label{15}
    C(t)=max(0,\sqrt{\lambda_{1}}-\sqrt{\lambda_{2}}-\sqrt{\lambda_{3}}-\sqrt{\lambda_{4}})
\end{equation}
where $\lambda'$s are the eigenvalues of the non-hermitian matrix
$\rho(t)\widetilde{\rho}(t)$ arranged in decreasing order of the
magnitude. The matrix $\rho(t)$ is the density matrix for the
two-bipartite and the matrix $\widetilde{\rho}(t)$ is given by
\begin{equation}\label{16}
    \widetilde{\rho}(t)=(\sigma_{y}^{A} \otimes\sigma_{y}^{B} )\rho^{*}(t) (\sigma_{y}^{A}\otimes
    \sigma_{y}^{B})
\end{equation}
where $\rho(t)^{*}$ is the complex conjugation of $\rho (t)$ and
$\sigma_{y}$ is the Pauli matrix given in quantum mechanics. The
concurrence fluctuated between $C=0$ for a separable state and
$C=1$ for a maximally entangled state. The two measures of
entanglement are different for mixed quantum states.\\
Here we will consider some interesting initial entangled states
for the two-bipartite which can be prepared and have potential
applications in the quantum information processing tasks
\cite{BBCJPW93,BPMEWZ97,BBDHP98,FSBFKP98,BW92,E91}.\\
We will start by the investigation of the EPR-states which are
concepts in quantum information science, a crucial part of quantum
teleportation and represent the simplest possible examples of
entanglement.
\begin{enumerate}
\item Assume that the initially entangled state of the field in two cavities to be in a
NOON state given by
\begin{equation}\label{151}
|\Psi\rangle_{AB}(0)=a_{2}|0\rangle_{A}|1\rangle_{B}+
    a_{3}|1\rangle_{A}|0\rangle_{B}
\end{equation}
This kind of state can be generated by passing a two-level atom
initially in the excited state through the two empty high-Q
cavities. The interaction times of an atom with two cavities are
chosen to be such that we have a $\pi/2$ pulse in the first cavity
and a $\pi$ pulse in the second cavity \cite{DZBRH94}. The initial
logarithmic negativity and concurrence are given by
\begin{equation}\label{24}
    \mathcal{N}(0)=\log_{2}(1+2|\rho_{23}(0)|)=\log_{2}(1+2|a_{2}a_{3}|)
\end{equation}
\begin{equation}\label{25}
    C(0)=2|\rho_{23}(0)|=2|a_{2}a_{3}|
\end{equation}
where the two quantities are related for this case by
$\mathcal{N}(0)=\log_{2}(1+C(0))$. It is evident, the solution of
this equation gives $\mathcal{N}(0)=C(0)=1$ which corresponds to
the case $|a_{2}|=|a_{3}|=\frac{1}{\sqrt{2}}$. In (Figure
\ref{Fig1}) the variation of the initial values of the logarithmic
negativity and concurrence in terms of $|a_{2}|$ is presented
where
$|a_{2}|^{2}+|a_{3}|^{2}=1$. \\
Using the solutions of Appendix A, it can be shown that the
density matrix $\rho(t)$ can be read as
\begin{equation}\label{21}
    \rho(t)=\left(%
\begin{array}{cccc}
  \rho_{11}(t) & 0 & 0 & 0 \\
  0 & \rho_{22}(t) & \rho_{23}(t) & 0 \\
  0 & \rho_{23}^{*}(t) & \rho_{33}(t)& 0 \\
  0 & 0 & 0 & \rho_{44}(t) \\
\end{array}%
\right)
\end{equation}
Then, we can calculate the negativity defined by (\ref{201}) for
the two-bipartite. We find
\begin{eqnarray}
\mathcal{N}(t)&=& max \bigg(0, \nonumber \\
&&\log_{2}\bigg[1-\rho_{11}(t)-\rho_{44}(t)+\sqrt{[\rho_{11}(t)-\rho_{44}(t)]^{2}+4|\rho_{23}(t)|^{2}}\bigg]\bigg)\nonumber
\\\label{22}
\end{eqnarray}
while the concurrence defined by (\ref{15}) is given by
\begin{equation}\label{23}
    C(t)=max\left(0, 2\left[|\rho_{23}(t)|-\sqrt{\rho_{11}(t)\rho_{44}(t)} \right]\right)
\end{equation}
As they decay, they get entangled with the environment, slowly
losing their coherence and purity over time. \\
For the case of vacuum reservoir, we can see from the solutions of
the Appendix A that $\rho_{44}(t)=0$. Then the relation between
concurrence and logarithmic negativity can be given by
\begin{eqnarray}
\mathcal{N}(t)&=& max \bigg(0,\log_{2}\bigg[1-\rho_{11}(t)+
\sqrt{\rho_{11}(t)^{2}+C^{2}}\bigg]\bigg)\label{201}\\
C&=&2|\rho_{23}(t)|\label{204}
\end{eqnarray}
where we can not observe in this case the ESD for any initial
states.

\item Consider now the initially entangled two-bipartite to be in a
another EPR-state given by
\begin{equation}\label{26}
    |\Psi\rangle_{AB}(0)=a_{1}|0\rangle_{A}|0\rangle_{B}+a_{4}|1\rangle_{A}|1\rangle_{B}
\end{equation}
This kind of states can be prepared by swapping the vacuum and
one-photon state in the cavity A of the state (\ref{151})
discussed above \cite{S06,IZZ00}. States like this have been
realized in experiments with trapped ions \cite{L03,R04}. The
initial logarithmic negativity and concurrence read as
\begin{equation}\label{30}
    \mathcal{N}(0)=\log_{2}(1+2|\rho_{14}(0)|)=\log_{2}(1+2|a_{1}a_{4}|)
\end{equation}
\begin{equation}\label{31}
    C(0)=2|\rho_{14}(0)|=2|a_{1}a_{4}|
\end{equation}
where the maximum value of $\mathcal{N}(0)$ and $C(0)$ are equal 1
for $|a_{1}|=|a_{4}|=\frac{1}{\sqrt{2}}$, where
$|a_{1}|^{2}+|a_{4}|^{2}=1$. The initial values of logarithmic
negativity and concurrence have the same behavior (Figure
\ref{Fig1}) in terms of $|a_{1}|$.\\
Exploiting the solutions of the Appendix A, the density matrix
$\rho(t)$ can have the form
\begin{equation}\label{27}
    \rho(t)=\left(%
\begin{array}{cccc}
  \rho_{11}(t) & 0 & 0 & \rho_{14}(t) \\
  0 & \rho_{22}(t) & 0 & 0 \\
  0 & 0 & \rho_{33}(t)& 0 \\
  \rho_{14}^{*}(t) & 0 & 0 & \rho_{44}(t) \\
\end{array}%
\right)
\end{equation}
We can show that the logarithmic negativity defined by
(\ref{14})and the concurrence defined by (\ref{15}) for the
two-bipartite in this case are given by
\begin{eqnarray}
\mathcal{N}&(t)&=max\bigg(0, \nonumber \\
  & &
  \log_{2}\bigg[1-\rho_{22}(t)-\rho_{33}(t)+\sqrt{[\rho_{22}(t)-\rho_{33}(t)]^{2}+
  4|\rho_{14}(t)|^{2}}\bigg]\bigg)\nonumber \\\label{28}
\end{eqnarray}
and
\begin{equation}\label{29}
    C(t)=max\left(0, 2\left[|\rho_{14}(t)|-\sqrt{\rho_{22}(t)\rho_{33}(t)} \right]\right)
\end{equation}
As is evident form the solutions in Appendix A, in this case
$\rho_{22}(t)=\rho_{33}(t)$. Then the relation between concurrence
and logarithmic negativity can be given by
\begin{eqnarray}\label{202}
\mathcal{N}(t)&=& max \bigg(0,\log_{2}[1+\widetilde{C}(t)]\bigg)
\\
C(t)&=& max(0,\widetilde{C}(t))
\end{eqnarray}
where\begin{equation}\label{203}
    \widetilde{C}(t)=2\left[|\rho_{14}(t)|-\rho_{22}(t)\right]
\end{equation}
which makes clear that the logarithmic negativity is greater then
the concurrence, when $C$ decreases in terms of $t$, except for
the initial value which is one for both of them for
$|a_{1}|=|a_{4}|=\frac{1}{\sqrt{2}}$, and the value for
$t\rightarrow \infty$ which is also zero.
\end{enumerate}

\section{Entanglement dynamics}\label{sec4}
\begin{figure}
  \center{
  \includegraphics[height=0.4\textheight]{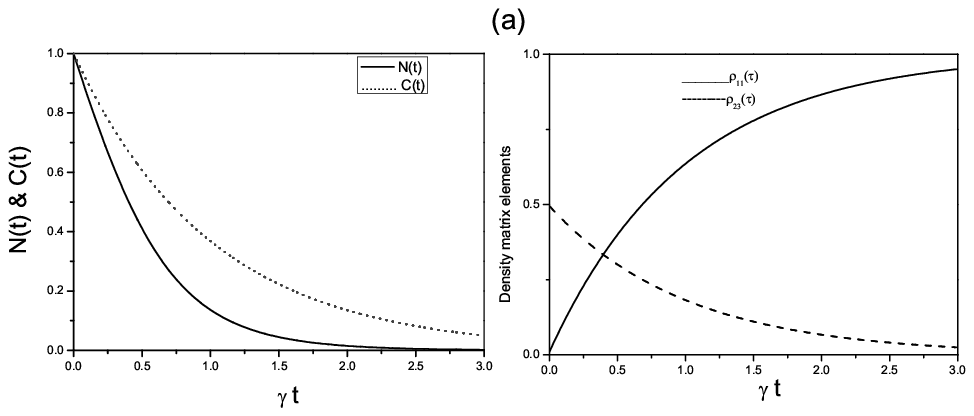}\\
  \includegraphics[height=0.4\textheight]{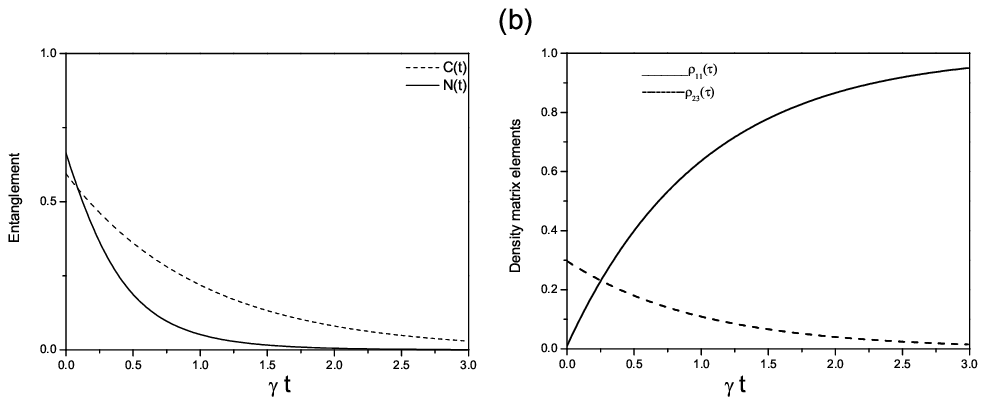}\\}
  \caption{Entanglement dynamics of the first EPR state for different
  initial probability for vacuum reservoirs.
 (a) $|a_{2}|=\frac{1}{\sqrt{10}}$, (b) $|a_{2}|=\frac{1}{\sqrt{2}}$.
    }\label{Fig3}
\end{figure}

\begin{figure}
  \center{
  \includegraphics[height=0.4\textheight]{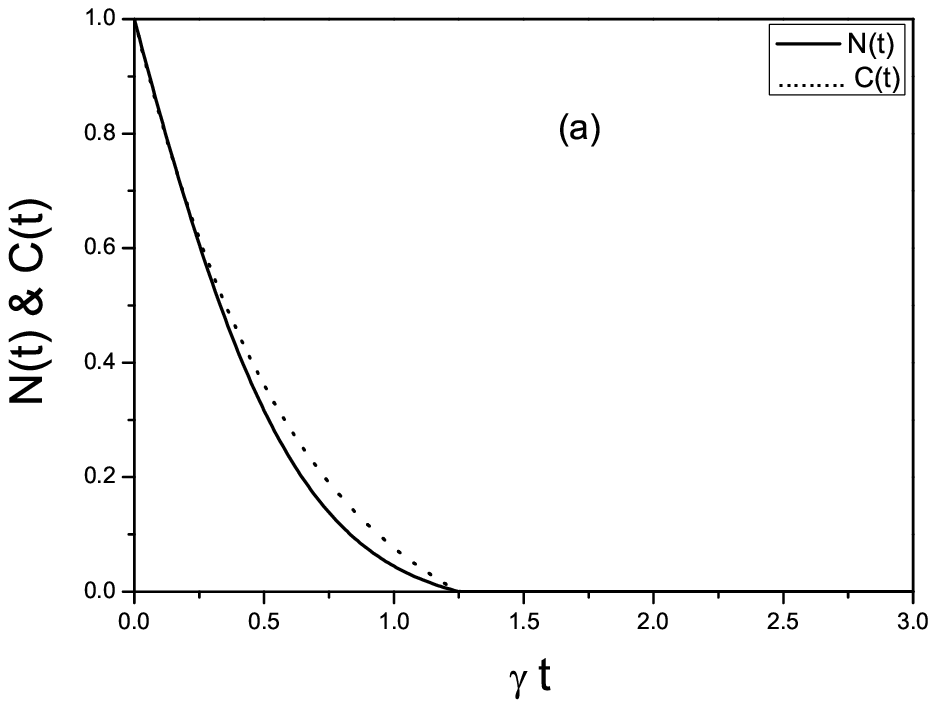}\\
  \includegraphics[height=0.4\textheight]{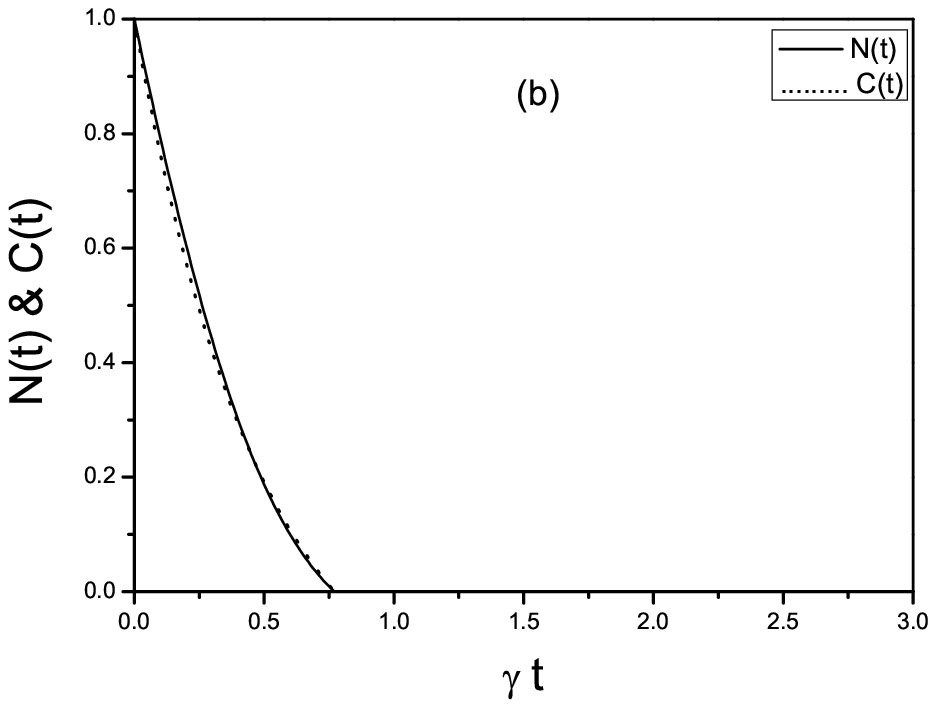}}\\
  \caption{Entanglement dynamics of the first EPR state at
  different mean photon numbers of reservoirs
  in the case
  $|a_{2}|=|a_{3}|=\frac{1}{\sqrt{2}}$.
   (a)  $\overline{n}=0.1$, (b) $\overline{n}=0.25$. }\label{Fig31}
\end{figure}

\begin{figure}
  \center{
  \includegraphics[height=0.45\textwidth]{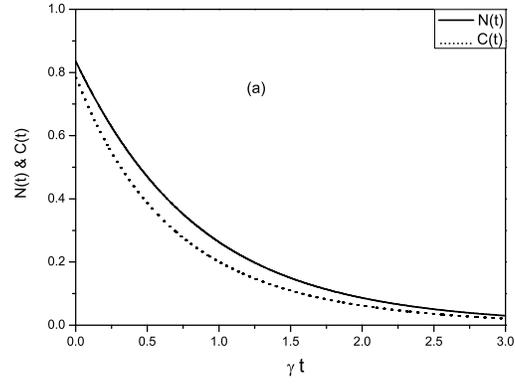}\\
   \includegraphics[height=0.45\textwidth]{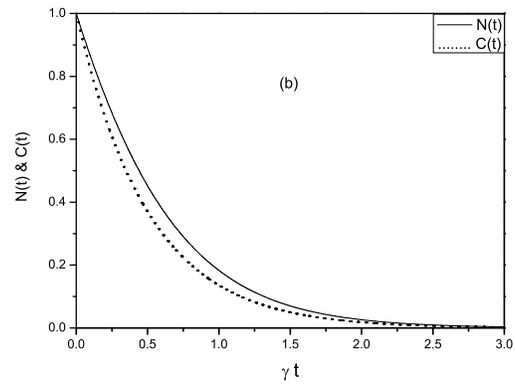}\\
  \includegraphics[height=0.45\textwidth]{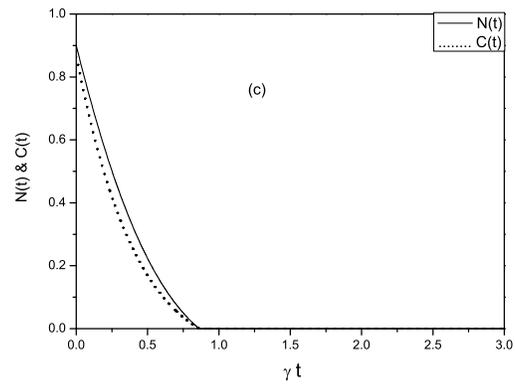}}
  \caption{Entanglement dynamics of the second EPR state at different initial probability for vacuum reservoirs.
   (a) $|a_{1}|=0.9$, (b) $|a_{1}|=\frac{1}{\sqrt{2}}$, (c) $|a_{1}|=0.5$. }\label{Fig4}
\end{figure}

\begin{figure}
  \center{
  \includegraphics[height=0.4\textheight]{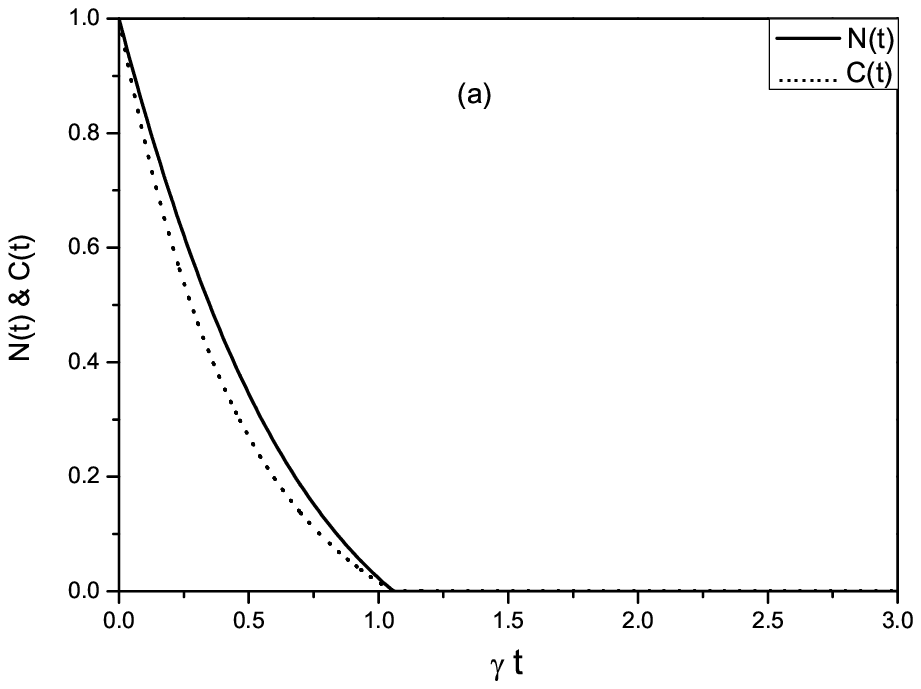}\\
  \includegraphics[height=0.4\textheight]{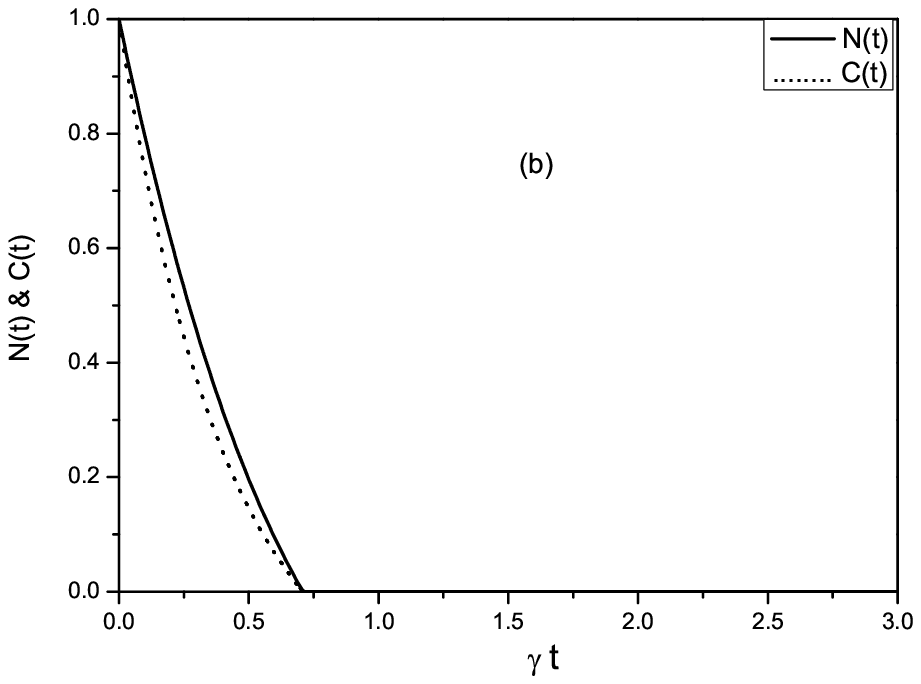}}
  \caption{Entanglement dynamics of the second EPR state at different mean photon numbers of
  reservoirs in the case $|a_{1}|=|a_{4}|=\frac{1}{\sqrt{2}}$
   (a) $\overline{n}=0.1$, (b) $\overline{n}=0.25$. }\label{Fig5}
\end{figure}

We will explore the time evolution of entanglement of the previous
entangled states of two-bipartite system exposed to either vacuum
or thermal reservoirs. In figure \ref{Fig3} we plot the
logarithmic negativity and the concurrence for the first EPR state
discussed previously in (i), for different initial states for
vacuum reservoirs.  Note that for vacuum reservoirs
$\overline{n}=0$, i.e. at zero temperature , no ESD is observed.
$\mathcal{N}(t)$ and $C(t)$ monotonically decrease to zero as
$t\rightarrow\infty$. On the other hand, the two measures give
different values for entanglement in vacuum reservoirs. Generally,
the logarithmic negativity takes smaller values than the
concurrence \cite{TF04}, except for initial value for some initial
states figure (\ref{Fig3},b), this can be seen from the equations
(\ref{201}, \ref{204})where the concurrence is proportional to the
population $\rho_{23}(t)$ which monotonically decreases, while the
logarithmic negativity depends on the population $\rho_{11}(t)$
which growths rapidly to rich a maximum value but the population
$\rho_{44}(t)$ remains zero. In figure \ref{Fig31} we present the
two measures of entanglement for the thermal reservoirs in the
case $|a_{2}|=|a_{3}|=\frac{1}{\sqrt{2}}$ and for different values
of $\overline{n}$. Here the ESD is observed and as the temperature
increases, the sudden death time decreases. On the other hand, we
can observe the perfect correspondence between the logarithmic
negativity and the concurrence for the thermal reservoirs while in
the vacuum reservoirs the concurrence exceed the logarithmic
negativity which decays faster than the concurrence where The
population $\rho_{44}(t)$ starts to manifest which restrains the
increasing population $\rho_{11}$. The sudden death time is
identical with the two measures of entanglement. Indeed, for the 2
two-dimensional systems, logarithmic negativity and concurrence
are real entanglement measures: they do not increase local
operations and classical communication, and vanish if only if the
state is separable.\\
In figure \ref{Fig4} we plot the dynamical evolution of the
entanglement when the second EPR state (ii) is considered, for
$|a_{1}|=|a_{4}|=\frac{1}{\sqrt{2}}$ and for different values of
$\overline{n}$. For this case no ESD in the vacuum reservoirs, but
when $a_{1}<a_{4}$ we can see a finite-time disentanglement, which
means that the major contribution of this state is responsible for
the ESD. Thus, we can conclude that locally equivalent pure states
with the same entanglement perform very differently during the
time evolution and simple local unitary operation acts on the
initial state can give naissance to ESD. As the mean photon
numbers in cavities increase, the ESD is always observed and the
finite-time disentanglement persists as shown in figure
\ref{Fig5}. Contrary to the first EPR state, the logarithmic
negativity exceeds the concurrence and they are in good
coincidence. Furthermore, the sudden death time is the same with
the two measures of entanglement.\\
On the other hand, it is obvious that the two measures of
entanglement are very closed. The sudden death time is the same in
the two measures of entanglement, and the logarithmic negativity
predicts the same behavior of the entanglement as the concurrence.
When the average thermal photon number is different from zero, we
can see that the ESD always occurs whatever the initial entangled
states are and no matter how the nonzero average thermal photon
number is. Furthermore, the two proposed measures of entanglement
coincide and give a good prediction of the sudden death
entanglement of the two-bipartite. This is consistent with the
findings in \cite{ILZ07, TIAZ08}

\section{Conclusion}\label{sec5}
The delicate aspect of entanglement has been subjected to many
quantitative studies and has guided to interesting results.
Understanding the physical meaning of entanglement measures
continues, however, a major defy. In this work we have shown that
the scheme we propose here allows direct measures of entanglement:
logarithmic negativity and concurrence of a two-bipartite cavity
system. The sudden death time depends on the number of photons in
the cavities and the temperature of the reservoirs. It increases
with increasing the number of photons in the cavities and
decreases with increasing temperature of the reservoirs. The
proposed scheme only involves system-reservoir interactions
corresponding to the decay of the radiation field inside a cavity
(field damping). This operation is based on the elementary
exchange of energy between system and reservoir that is thus
assumed to consist of the simultaneous creation of a quantum
excitation of the system with annihilation of a quantum in one
mode of the reservoir, or reverse process. These operations have
been demonstrated experimentally \cite{BORNABRH01} and therefore
our proposed scheme can be realized within the present cavity-QED
technologies. As a future perspective il would be interesting to
study the effect of other type of environment like a squeezed
reservoirs. Further, It would be motivating to enlarge our point
of view to the two-bipartite entanglement of high dimension for
the study disentanglement and loss of decoherence, where the
Wootter's criterion is not applicable and we can use logarithmic
negativity which is necessary and sufficient condition for $2 X 3$
and $3 X 2$ systems.

\appendix
\section{Equations of motion of the density matrix elements and their
solutions for vacuum reservoir}

The equation of motion of density matrix elements for the general
state (Eq. \ref{19}) are given by
\begin{eqnarray*}
\dot{\rho}_{11}(t)&=&
(-\overline{n}_{A}\gamma^{(A)}-\overline{n}_{B}\gamma^{(B)})\rho_{11}(t)+
(\overline{n}_{A}+)\gamma^{(A)}\rho_{33}(t)+(\overline{n}_{B}+1)\gamma^{(B)}\rho_{22}(t)\\
\dot{\rho}_{12}(t) &=&
-[\frac{1}{2}(2\overline{n}_{B}+1)\gamma^{(B)}+\overline{n}_{A}\gamma^{(A)}]\rho_{12}(t)+(\overline{n}_{A}+1)\gamma^{(A)}\rho_{34}(t)\\
\dot{\rho}_{13}(t)&=&
-[\frac{1}{2}(2\overline{n}_{A}+1)\gamma^{(A)}+\overline{n}_{B}\gamma^{(B)}]\rho_{13}(t)+(\overline{n}_{B}+1)\gamma^{(B)}\rho_{24}(t)\\
\dot{\rho}_{14}(t)&=&
-[(\overline{n}_{A}+\frac{1}{2})\gamma^{(A)}+\gamma^{(B)}(\overline{n}_{B}+\frac{1}{2})]\rho_{14}(t)\\
\dot{\rho}_{21}(t) &=&
-[(\frac{1}{2}(2\overline{n}_{B}+1))\gamma^{(B)}+\overline{n}_{A}\gamma^{(A)}]\rho_{21}(t)+(\overline{n}_{A}+1)\gamma^{(A)}\rho_{43}(t)\\
\dot{\rho}_{22}(t)&=&
-[\overline{n}_{A}\gamma^{(A)}+(\overline{n}_{B}+1)\gamma^{(B)}]\rho_{22}(t)+
(\overline{n}_{A}+1)\gamma^{(A)}\rho_{44}(t)+\overline{n}_{B}\gamma^{(B)}\rho_{11}(t)\\
\dot{\rho}_{23}(t) &=&
-[(\overline{n}_{A}+\frac{1}{2})\gamma^{(A)}+(\overline{n}_{B}+\frac{1}{2})\gamma^{(B)}]\rho_{23}(t)\\
\dot{\rho}_{24}(t) &=&
-[(\overline{n}_{B}+1))\gamma^{(B)}+(\overline{n}_{A}+\frac{1}{2})\gamma^{(A)}]\rho_{24}(t)+\overline{n}_{B}\gamma^{(B)}\rho_{13}(t)\\
\dot{\rho}_{31}(t) &=&
-[\frac{1}{2}(2\overline{n}_{A}+1))\gamma^{(A)}+\overline{n}_{B}\gamma^{(B)}]\rho_{31}(t)+(\overline{n}_{B}+1)\gamma^{(B)}\rho_{42}(t)\\
\dot{\rho}_{32}(t) &=&
-[(\overline{n}_{A}+\frac{1}{2})\gamma^{(A)}+(\overline{n}_{B}+\frac{1}{2})\gamma^{(B)}]\rho_{32}(t)\\
\dot{\rho}_{33}(t) &=&
-[(\overline{n}_{A}+1)\gamma^{(A)}+\overline{n}_{B}\gamma^{(B)}]\rho_{33}(t)+
\overline{n}_{A}\gamma^{(A)}\rho_{11}(t)+(\overline{n}_{B}+1)\gamma^{(B)}\rho_{44}(t)\\
\dot{\rho}_{34}(t) &=&
-[(\overline{n}_{A}+1)\gamma^{(A)}+(\overline{n}_{B}+\frac{1}{2})\gamma^{(B)}]\rho_{34}(t)+
\overline{n}_{A}\gamma^{(A)}\rho_{12}(t)\\
\dot{\rho}_{41}(t) &=&
-[(\overline{n}_{A}+\frac{1}{2})\gamma^{(A)}+(\overline{n}_{B}+\frac{1}{2})\gamma^{(B)}]\rho_{41}(t)\\
\dot{\rho}_{42}(t) &=&
-[(\overline{n}_{A}+\frac{1}{2})\gamma^{(A)}+(\overline{n}_{B}+1))\gamma^{(B)}]\rho_{42}(t)+
\overline{n}_{B}\gamma^{(B)}\rho_{31}(t)\\
\dot{\rho}_{43}(t) &=&
-[(\overline{n}_{A}+1)\gamma^{(A)}+(\overline{n}_{B}+
\frac{1}{2})\gamma^{(B)}]\rho_{43}(t)+\overline{n}_{A}\gamma^{(A)}\rho_{21}(t)\\
\dot{\rho}_{44}(t) &=&
-[(\overline{n}_{A}+1)\gamma^{(A)}+(\overline{n}_{B}+1)\gamma^{(B)}]\rho_{44}(t)+
\overline{n}_{A}\gamma^{(A)}\rho_{22}(t)+\overline{n}_{B}\gamma^{(B)}\rho_{33}(t)
\end{eqnarray*}
For the sake of simplicity, we assume that the cavities are
identical $\gamma^{(A)}=\gamma^{(B)}=\gamma$ and
$\overline{n}_{A}=\overline{n}_{B}=\overline{n}$. On solving these
equations we get the time evolution of the density elements matrix
\begin{eqnarray*}
\rho_{11}(t) &=& \frac{1}{4a^{2}}
\{[-2+2a(2\rho_{11}(0)+\rho_{22}(0)+\rho_{33}(0)-1)](1+a)e^{-\eta
t}\nonumber\\& +&  [1+a^{2}(1-2\rho_{33}(0)-2\rho_{22}(0))\nonumber \\
&-&
  2a(2\rho_{11}(0)+\rho_{22}(0)+\rho_{33}(0)-1)]e^{-2\eta t}\nonumber\\
  &+&(1+a)^{2}\}\label{3}\\
\rho_{22}(t) &=& \frac{1}{4a^{2}}
\{[-1+a^{2}(2\rho_{22}(0)+2\rho_{33}(0)-1)\nonumber\\
&+&2a(2\rho_{11}(0)+\rho_{22}(0)+\rho_{33}(0)-1)]
e^{-2\eta t}\nonumber\\
&+&[2+2a^{2}(\rho_{22}(0)-\rho_{33}(0))-2a(2\rho_{11}(0)+\rho_{22}(0)+\rho_{33}(0)-1))
]e^{-\eta t}\nonumber\\
&+&(a^{2}-1)\}\label{4}\\
\rho_{33}(t) &=&\frac{1}{4a^{2}}
\{[-1+a^{2}(2\rho_{22}(0)+2\rho_{33}(0)-1)\nonumber\\
&+&2a(2\rho_{11}(0)+\rho_{22}(0)+\rho_{33}(0)-1)]e^{-2\eta t}\nonumber\\
&+&[2+2a^{2}(\rho_{33}(0)-\rho_{22}(0))-2a(2\rho_{11}(0)+\rho_{22}(0)+\rho_{33}(0)-1)]e^{-\eta t}\nonumber\\
&+&(a^2-1)\}\label{5}\\
\rho_{44}(t) &=&1-\rho_{11}-\rho_{22}-\rho_{33}\label{6}\\
\rho_{12}(t) &=&\frac{1}{2a}
\{[a(\rho_{12}(0)-\rho_{34}(0))-\rho_{34}(0)-\rho_{12}(0)]e^{-\frac{3}{2}\eta
t}\nonumber\\
&+&(\rho_{34}(0)+\rho_{12}(0))(1+a)e^{-\frac{1}{2}\eta t}\}\label{7}\\
\rho_{13}(t) &=&\frac{1}{2a}
\{[a(-\rho_{24}(0)+\rho_{13}(0))-\rho_{24}(0)-\rho_{13}(0)]e^{-\frac{3}{2}\eta
t}\nonumber\\&+&(\rho_{24}(0)+\rho_{13}(0))(1+a))e^{-\frac{1}{2}\eta
t}\}\label{8}\\
\rho_{14}(t) &=& \rho_{14}(0)e^{-\eta t}\label{9}\\
\rho_{23}(t) &=& \rho_{23}(0)e^{-\eta t}\label{10}\\
\rho_{24}(t) &=&\frac{1}{2a}
\{(a(\rho_{24}(0)-\rho_{13}(0))+\rho_{24}(0)+\rho_{13}(0))e^{-\frac{3}{2}\eta
t}\nonumber\\
&+&(\rho_{24}(0)+\rho_{13}(0))(-1+a))e^{-\frac{1}{2}\eta t}\}\label{11}\\
\rho_{34}(t) &=&\frac{1}{2a}
\{[a(\rho_{34}(0)-\rho_{12}(0))+\rho_{34}(0)+\rho_{12}(0)]e^{-\frac{3}{2}\eta t}\nonumber\\
&+&(\rho_{34}(0)+\rho_{12}(0))(-1+a)e^{-\frac{1}{2}\eta
t}\}\label{12}
\end{eqnarray*}
and $\rho_{21}(t)=\rho_{12}^{}(t), \rho_{31}(t)=\rho_{13}^{}(t),
\rho_{32}(t)=\rho_{23}^{}(t), \rho_{41}(t)=\rho_{14}^{}(t),
\rho_{42}(t)=\rho_{24}^{}(t), \rho_{43}(t)=\rho_{34}^{}(t)$, where
$a=2\overline{n}+1$ and $\eta=a\gamma$.

\newpage

\end{document}